\documentclass[aps,prl,twocolumn]{revtex4}
\usepackage[usenames]{color}
\newcommand{\comment}[1]{}
\usepackage{graphicx}
\usepackage{amssymb}
\usepackage{epsfig}
\newcommand{\bra}[1]{\langle {#1} |}
\newcommand{\ket}[1]{| {#1} \rangle}

\newcommand{\ketn}[1]{ {#1} \rangle}

\begin{document}

\bibliographystyle{revtex}
\title{Reply to comment by S.-K. Yip cond-mat/0611426 }
\author{Ryan Barnett$^1$, Ari Turner$^2$, and
Eugene Demler$^2$}
\affiliation{$^1$Department of Physics, California Institute of Technology, MC 114-36, Pasadena, California 91125,USA}
\affiliation{$^2$Department of Physics, Harvard University, Cambridge, Massachusetts 02138, USA}
\date{\today}
\begin{abstract}
We respond to S.-K. Yip's criticism of our work on the
classification of spinor condensates. We explain why his criticism
is unfounded, emphasizing that the phases he mentions
have been addressed in our paper
cond-mat/0611230. To provide a constructive aspect to this response,
we use it as an opportunity to show how our classification
scheme makes explicit not only spin rotations which leave spinor
states invariant, but also the phase factors which need to accompany them.
\end{abstract}

\maketitle

In a comment \cite{yip06} S.-K. Yip criticized our recent work on the classification of spinor condensates \cite{barnett06a} and claimed that when discussing
residual symmetries of various ground states, we did not understand the importance of phase factors and their role in determining the nature of vortex
excitations.  This criticism is unfounded. Since the focus of \cite{barnett06a}
was on classifying the symmetries of the phases, with topological excitations
presented as an application, it was unnecessary to give a thorough treatment
of such phases. We described only the spin
structure of the defects, which does give a complete classification of defects 
in the Mott insulating
 phase since there is no phase coherence. On the other hand,
in the follow-up paper 
\cite{barnett06b}, the focus was on vortices in condensates.  
We applied
our approach to classify spinors for $S=3$ condensates and
explicitly discussed such phase factors (see, for example, Eq. (8)). 
This is the information needed 
to determine the quantization rule for the phase change
of different vortices.
Properties of the isotropy group fully determine the fundamental group and
the nature of line defects \cite{mermin79}.
This paper therefore addresses the issues that Yip claims we disregarded.
This work was not cited in \cite{yip06}, although it had been
available on the archive before he posted his comment. 

To provide a constructive aspect to this response, we will use it as an
opportunity to illustrate  
that our classification scheme makes explicit
not only spin rotations which leave spinor states invariant but also
the phase factors which, as Yip emphasizes, need to accompany them. 
We will show that our
classification scheme allows one to calculate these phases as simple
geometrical factors related to properties of polyhedra representing
various spinor states.
 
The result (which will be proved elsewhere \cite{turner06}) is as follows.
Consider a spinor $\ket{\psi}$, and
a rotation $R=e^{-i {\bf F}\cdot \hat{\bf n}\alpha}$ which
is a symmetry of the set of the $2F$ points on the unit sphere consisting 
of coherent states $\ket{\zeta}$ such that  $\bra{\psi}\ketn{\zeta}=0$
(which we refer to as ``spin roots'').
Then $R\ket{\psi}$ has the same set of 
spin roots.   Since the spin roots determine $\ket{\psi}$ up to phase
we have that, 
\begin{equation}
R\ket{\psi}=e^{i\lambda}\ket{\psi}.
\end{equation}
where $\lambda$ is a real number  which may be 
determined by considering the axis of rotation 
$\hat{\bf n}$. Let $r$ be the multiplicity of the spin root which
 coincides with the direction of this axis ($r=0$ corresponds to 
no root on the axis). Then 
\begin{equation}
\label{phase}
\lambda=\alpha (r-F),
\label{eq:main}
\end{equation}
where $\alpha$ is the angle of the rotation (counterclockwise when viewed
facing from the end of $\hat{\bf n}$ towards the origin).

We will now proceed to apply Eq.~(\ref{phase}) to the examples that Yip
considers \cite{yip06}.  For the first example of the spin-two 
ferromagnetic state we have a continuous symmetry of rotations by
$\alpha$ about the $z$-axis.  The spin roots consist of four
degenerate points on the negative $z$-axis.  Then for arbitrary
rotation by $\alpha$ counterclockwise about the positive $z$-axis we 
accumulate the phase $\lambda=\alpha(0-2)=-2\alpha$ (as could have
been obtained directly by applying the rotation $R$ to the
spinor).  
The second example is the spin-two 
tetrahedratic (i.e. cyclic) state which
has the spin
roots residing at the vertices of a regular tetrahedron.  This
will have as one of its symmetries a rotation by $2\pi/3$ about
an axis going through any of the vertices.  Such a rotation will
result in the phase $\lambda=\frac{2\pi}{3}(1-2)=-\frac{2\pi}{3}$.
As the last example, we take the (spin-three) hexagonal state.
This state has a symmetry of a rotation by $2\pi/6$ which
results in the phase  $\lambda=\frac{2\pi}{6}(0-3)=-\pi$.


\end{document}